\documentstyle[11pt,epsf]{article}
\title{
  {\vspace{-3cm} \normalsize \hfill
    \parbox{38mm}{
HD-THEP-00-02 \\       
WUE-ITP-2000-007 \\       
}}\\[25mm]
Learning structured data from
unspecific reinforcement  }

\author{
Michael Biehl$^1$,
Reimer K\"uhn$^2$,
Ion-Olimpiu Stamatescu$^{2,3}$\\[5mm]
{\small $^1$Institut f\"ur Theoretische Physik, Universit\"at W\"urzburg,
Am Hubland}\\ 
{\small D-97074 W\"urzburg, Germany}\\
{\small $^2$Institut f\"ur Theoretische Physik, Universit\"at Heidelberg,
 Philosophenweg 16 und 19}\\
{\small D-69120  Heidelberg, Germany}\\
{\small $^3$Forschungst\"atte der Evangelischen Studiengemeinschaft, 
Schmeilweg 5,}\\ 
{\small D-69118  Heidelberg, Germany}
}

\date{\today}

\newcommand{\be}{\begin{equation}}
\newcommand{\ee}{\end{equation}}
\newcommand{\bea}{\begin{eqnarray}}
\newcommand{\eea}{\end{eqnarray}}
\newcommand{\no}{\noindent}
\newcommand{\sign}{{\rm sign}}
\newcommand{\e}{{\rm e}}

\font\tams                   = cmmib10
\font\kleinhalbcurs          = cmmib10 scaled 833
\font\bxf                    = cmbx10
\font\sevenbf                = cmbx7 scaled \magstep1
\def\vec#1{{\textfont1=\tams\scriptfont1=\kleinhalbcurs
\textfont0=\bxf\scriptfont0=\sevenbf
\mathchoice
{\hbox{$\displaystyle#1$}}
{\hbox{$\textstyle#1$}}
{\hbox{$\scriptstyle#1$}}
{\hbox{$\scriptscriptstyle#1$}}}}

\begin{document}
\maketitle

\begin{abstract}
We show that a straightforward extension of a 
simple learning model based on the Hebb rule, the 
previously introduced
Association-Reinforcement-Hebb-Rule, can cope with
``delayed", unspecific reinforcement also in the case of structured data and lead to perfect generalization.

\end{abstract}

\section{Introduction}

Learning from unspecific reinforcement may be essential in various contexts,
both natural and artificial, where, typically, the results of
particular actions add to a final consequence which only is valuated.
The freedom residing in each step is not (or only partially) controlled
directly and the learner must cope with the necessity of improving its
performance only from information concerning the final success of a
complex series of actions.

It is therefore important to find out whether there are {\it simple and robust
procedures} for such situations, {\it which
might have developed under natural conditions and which
may be basic also for artificial learning rules} (for this reason we  
do not consider evolved algorithms like 
 Q-learning \cite{wat}, TD learning \cite{sut} etc). In previous 
works \cite{MS85},\cite{renu99} we have
introduced an ``Association-Reinforcement" learning model based on  the
following conception:\par
 1. For each given input (external situation) the agent answers with
an action (operation) depending solely on the input 
and on its instantaneous internal
(cognitive) structure and simultaneously strengthens 
(in its internal structure) the
 {\it blind} association between
this particular input and action.\par
 2. At the end of a series of actions (path) the final success is judged.
Then
the associations ``situation - operation" which have been involved on 
this path
are re-weighted equally and depending only on the final success -- 
{\it unspecific}
reinforcement.

In \cite{renu99} we studied an implementation of this model to 
a classification problem for perceptrons,
the
{\it Association-Reinforcement-Hebb-rule} and showed some 
amazing properties:\par
 a) Despite the fact that feedback on the
learner's performance enters its learning dynamics only in an {\em
unspecific\/} way in that it cannot be associated with single identifiable
correct or incorrect associations, convergence of the AR-Hebb-algorithm  in
the sense of {\em asymptotically perfect generalization\/} is found.\par
 b) For given initial conditions, this convergence depends on the
learning parameters characterizing the 2 steps described above;
in particular none of these steps  can be completely
inhibited. Alternatively, for given 
algorithm parameters convergence may depend
on initial conditions.

In detail the dynamics of this algorithm was found to be very complex and
interesting, being controlled by fixed points in the pre-asymptotic regime,
and having a continuous set of asymptotic convergence laws. These results could 
easily be extended to the more realistic case where in the second step 
the unspecific reinforcement is randomly applied to only part of the 
associations achieved in the first step (the agent does not recall 
everything it has done on the trial) \cite{renu99}. Further interesting 
extensions concern the question of structured data and of multi-layer 
perceptrons.

Structured data represent a more
involved classification problem and it is known that when 
teacher and  data vector are not fully aligned (or exactly
uncorrelated) the
usual Hebb rule does not lead to convergence of the student vector
onto that of the teacher,
while the perceptron algorithm does \cite{ribi}. On the other hand,
the limiting case of the AR-Hebb-rule corresponding to the perceptron rule
has been shown not to converge in the case of unspecific reinforcement
for non-structured data.
It is therefore a non-trivial question whether the unspecific reinforcement
problem can be solved for structured data and in particular, whether 
some immediate extension of the AR-Hebb-rule can be shown to converge
in this case. It is this question which we shall address in this paper.
In a future publication we shall treat the problem of the committee 
machine as a first step to multi-layer perceptrons.

In section 2 we shall describe the learning model in the general setting
and in section 3 we shall discuss its convergence properties
providing  numerical and analytical results. Thereby we shall briefly recall
the non-structured data case and then concentrate on the general,
structured data case. Section 4 contains the conclusions.

\section{Learning rule for perceptrons under unspecific reinforcement}

We consider one layer perceptrons with Ising or
real number units $s_i$,
 real weights (synapses) $J_i$ and one Ising output unit:
\begin{equation}
     s = \sign \left( \frac{1}{\sqrt{N}}\sum_{i=1}^N J_i\, s_i\right)
\label{e.1}
\end{equation}
\no Here $N$ is the number of input nodes, and we put no explicit thresholds.
The network (student) is presented with a series of patterns
$s_i=\xi_i^{(q,l)}$,  $q=1, ... ,Q$, $l=1, ... ,L$
to which it answers with $s^{(q,l)}$. A training period consists of
the successive presentation of $L$ patterns.
The answers are compared with the corresponding answers  $t^{(q,l)}$
of a teacher
with pre-given weights $B_i$ and
the average error made by the student over one
training period is calculated:
\be
    e_q = \frac{1}{2L} \sum_{l=1}^L |t^{(q,l)} -s^{(q,l)}| .
\label{e.2}
\ee
\no The training algorithm
consists of two parts:\par
\begin{itemize}
\item [I.] - a ``blind" Hebb-type  {\it association} at each presentation
of a pattern:
\be
   {J_i}^{(q,l+1)} ={J_i}^{(q,l)} + \frac{a_1}{\sqrt{N}}\, s^{(q,l)}\,
   {\xi}_i^{(q,l)};
\label{e.3}
\ee
\item [II.] - an ``unspecific" but graded {\it reinforcement} proportional
to the average error $e_q$ introduced in (\ref{e.2}), also Hebbian,
at the end of each training period,
\be
   {J_i}^{(q+1,1)} = {J}_i^{(q,L+1)} - \frac{a_2}{\sqrt{N}}\, e_q\,
                \sum_{l=1}^L r_l\, s^{(q,l)}\,{\xi}_i^{(q,l)}.
\label{e.4}
\ee
\no where $e_q$ is the average error eq. (\ref{e.2}) and $r_l$ is a 
dichotomic random variable: 
\be
r_l = \left\{ 
\begin{array}{lll}1 &{\rm with \ probability}& w \\
                 0& {\rm with \ probability}&1-w
\end{array} \right. 
\label{e.r}
\ee
\end{itemize}
\no Because of these 2 steps we  called this algorithm
 ``association/reinforcement(AR)-Hebb-rule".
We are
interested in the behavior with the number of iterations $q$ of 
the generalization
error $\epsilon_g(q)$:

\be
   \epsilon_g(q) = \frac{1}{\pi}{\rm arccos}
\left({{{\vec J}\cdot {\vec B}} \over {|{\vec J}|~|{\vec B}|}}\right),
\label{e.errg}
\ee

\no  in particular we  shall
test whether the behavior of $\epsilon_g(q)$
follows a power law at large $q$:
\be
      \epsilon_g(q) \simeq {\rm const}~ q^{-p}\ .
\label{e.6}
\ee

\no The training patterns $\{\xi_i^{(q,l)}\}$
are generated randomly from the following distribution:

\bea
P({\vec \xi}) &=& \frac{1}{2} \sum_{\sigma=\pm 1}P({\vec \xi}|\sigma) \nonumber \\
P({\vec \xi}|\sigma) &=& \prod_{i=1}^N \frac{1}{\sqrt{2\pi}}{\rm e}^{-\frac{1}{2}
(\xi_i - m \sigma C_i )^2} 
\label{e.dpbb}
\eea

\no and we take:

\be
{\vec C}^2 = {\vec B}^2 = N, \ \ {\vec C}\cdot {\vec B} = \eta N
\label{e.nor}
\ee

\no with fixed, given $m, \eta$. 
Notice the following features:
\begin{itemize}
\item [a)] During training the student only uses its own associations 
${\vec \xi}^{(q,l)}\leftrightarrow s^{(q,l)}$ 
and the average error $e_q$ which does not refer
specifically
 to the particular steps $l$.
\item [b)] Since the answers $s^{(q,l)}$ are made on the basis of the 
instantaneous
weight values ${\vec J}^{(q,l)}$ which change at 
each step according to eq. (3),
the series of
answers form a correlated sequence with each step depending on the previous 
one.
Therefore $e_q$ measures in fact the performance of a ``path",
an interdependent set of
 decisions.
\item[c)] In contrast with the case studied in \cite{renu99} the patterns 
can now have a structure. This introduces essential differences  to the 
previous situation, as we shall see in the next section.
\item [d)] We  explicitly account for imperfect 
recall  at the reinforcement step by
 the parameter $w$ (\ref{e.r}). This introduces a supplementary,
biologically motivated 
randomness which, as already suggested in \cite{renu99}, 
does not appear to introduce qualitative changes in
the results, however (see section 3).
\item [e)] For $L=1$ (and $w=1$) the algorithm reproduces the usual
``perceptron rule" (for
$a_1=0$) or to the usual ``unsupervised Hebb rule" (for $a_2 = 2 a_1$)
for on-line learning, for which the corresponding asymptotic behavior
is known \cite{biri}, \cite{val}, \cite{ribi}.
\end{itemize}

To study the learning behaviour we use Monte Carlo simulation and coarse 
grained analysis. The latter is provided by combining
the {\it blind association} (\ref{e.3}) during a learning period of $L$
elementary steps and the graded {\it unspecific reinforcement} (\ref{e.4}) at
the end of each learning period into one coarse grained step

\bea
{J}_i^{(q+1,1)} = {J}_i^{(q,1)} + \frac{1}{\sqrt{N}}(a_1-a_2 e_q)\sum_{l=
1}^L r_l\,
\sign \left( \frac{1}{\sqrt{N}}\sum_{j=1}^N J_j\,{\xi}_j^{(q,l)}\right)\, 
{\xi}_i^{(q,l)}, \label{e.cgu}\\
e_q = \frac{1}{2L}\sum_{l=1}^L \left|\,
\sign \left( \frac{1}{\sqrt{N}}\sum_{k=1}^N J_k\,
{\xi}_k^{(q,l)}\right) -\,
\sign \left( \frac{1}{\sqrt{N}}\sum_{i=k}^N B_k\,
{\xi}_k^{(q,l)}\right)\right|.\label{e.cge}
\eea

For simplicity we shall take for the time being 
 $r_l=1$, i.\,e. $w=1$ in eq. (\ref{e.r}). We
introduce 
\be
\alpha=qL/N,\ \ \lambda=a_1/a_2
\ee

\no and rescale everything with $a_2$, which means that we can
take without loss of generality $a_2=1$ in (\ref{e.cgu}),(\ref{e.cge}). 
We define the overlaps:

\be
{\cal R}(\alpha) = \frac{1}{N}{\vec B}\cdot{\vec J}^{(q,l)}\quad ,
 {\cal Q}(\alpha) = \frac{1}{N}[{\vec J}^{(q,l)}]^2 \quad , 
 {\cal D}(\alpha) = \frac{1}{N}{\vec C}\cdot{\vec J}^{(q,l)}\quad .
 \label{e.cgdef}
 \ee

\no Note that
in the ``thermodynamic limit" $L/N \rightarrow 0$ the overlaps
are self-averaging and we  can neglect the
dependence of ${\cal R}$, ${\cal D}$ and ${\cal Q}$ on $l$.
We shall follow standard procedures \cite{val}, \cite{her}, \cite{kica},
\cite{bisch}. Treating $\alpha$ as a continuous variable and using:
\bea
x &\equiv& \pi \epsilon_g = \arccos \left(\frac{\cal R}
{\sqrt{\cal Q}}\right), \label{e.x2}\\
y &\equiv& \arccos \left(\frac{\cal D}
{\sqrt{\cal Q}}\right), \label{e.y2}\\
z &\equiv& \arccos \eta
\eea
\no we obtain the
coarse grained equations:
\bea
\frac{d{\cal R}}{d\alpha} &=& \frac{d}{d\alpha}\left(\sqrt{{\cal Q}}\cos x\right) = 
\left(\lambda-\frac{1}{2}\right) A_{JT}+\frac{1}{2L} 
A_{TT}+\frac{1}{2}\left(1-\frac{1}{L}\right) S_{JT}A_{JT}, \label{e.dr} \\
\frac{d{\cal D}}{d\alpha} &=& \frac{d}{d\alpha}\left(\sqrt{{\cal Q}}\cos y\right) =
\left(\lambda-\frac{1}{2}\right) A_{JC}+\frac{1}{2L} 
A_{TC}+\frac{1}{2}\left(1-\frac{1}{L}\right) S_{JT}A_{JC}, \label{e.dd} \\
\frac{d\sqrt{{\cal Q}}}{d\alpha} &=& \left(\lambda-
\frac{1}{2}\right) A_{JJ}+\frac{1}{2L} 
A_{TJ}+\frac{1}{2}\left(1-\frac{1}{L}\right) S_{JT}A_{JJ} + \nonumber\\
& &\frac{1}{2\sqrt{{\cal Q}}}\left[\left(\lambda-\frac{1}{2}\right)S_{JT}+
\frac{1}{4}\left(1-\frac{1}{L}\right)S_{JT}^2 + 
\left(\lambda-\frac{1}{2}\right)^2 +\frac{1}{4L}\right], \label{e.dq2}
\eea
\no where the expectation values $A_., S_.$ are given in Appendix (section \ref{app1}).
These equations describe the flow of the three quantities 
$\epsilon_g = x/\pi $, ${\cal Q} $ and $y $ with $\alpha $ and involve the
data/teacher parameters $m$ and $z={\rm arcos}\eta $ and the learning parameter
$\lambda $. Note the geometric constraint:
\be
\sin\frac{y}{2} - \sin \frac{z}{2} = \omega\,  \sin \frac{x}{2}, \ \ 
|\omega| \le 1. \label{e.con}
\ee

\section{Convergence behaviour of the AR-Hebb algorithm}

\subsection{Non-structured data}

The case of non-structured data -- $m=0$ in eq. (\ref{e.ajt})-(\ref{e.erf}) -- has been
treated in \cite{renu99}, here we briefly recall some of the results for the 
later comparison with the structured data case.

Monte Carlo simulations indicate that in spite of the partial information
contained in the unspecific reinforcement
 perfect generalization is achieved by the
AR-Hebb algorithm and it depends on the learning parameters -- 
see \cite{renu99}. This intriguing behaviour is elucidated by the coarse grained
analysis. In this case eqs. (\ref{e.dr})-(\ref{e.dq2}) reduce to two equations
(for ${\cal R}$ and ${\cal Q}$) which have as general asymptotic solutions
 
\bea
\epsilon_g^2 &\simeq& \frac{1}{2\pi(\frac{1}{\lambda L}-1)}\, 
{\alpha}^{-1} + {\tilde c}_1 {\alpha}^{-\frac{1}{\lambda L}}\ \ \ \  \ 
{\rm for}\  
\lambda \neq \frac{1}{L}, \label{e.easa1}\\ 
\epsilon_g^2 &\simeq&\left(\frac{1}{2\pi} {\rm ln}\,{\alpha} + 
{\tilde c}_2\right) 
{\alpha}^{-1} \ \ \ \ \ 
{\rm for}\ \lambda = \frac{1}{L},  \label{e.easa2}\\ 
 {\cal Q} &\simeq& \frac{2}{\pi}\lambda^2\, \alpha^2 \label{e.qas} 
\eea 
 
\no at large $\alpha$. 
We see that for $\lambda < \frac{1}{L}$ we obtain asymptotically 
perfect generalization, 
the dominant term exhibiting  
 the usual power -1/2 , while for $\lambda > \frac{1}{L}$ the second term in 
(\ref{e.easa1}) dominates and   ensures again 
 perfect generalization but with a different power law, $-1/(2\lambda L)$. 
For $\lambda = \frac{1}{L}$ we obtain logarithmic corrections -- see eq.  
(\ref{e.easa2}). Notice that these results hold also for $L=1$.
One can generally see that for $\lambda = 0$ one cannot have perfect 
generalization for $L > 1$. For $L=1$ one re-obtains the asymptotic behavior found in 
\cite{biri}.
\par\bigskip

This learning algorithm
is further characterized by highly interesting pre-asymptotics,  
dominated by two stationarity conditions, one for the self-overlap, $d {\cal 
Q}/d \alpha = 0$, and one for the generalization error
$d\epsilon_g/d \alpha = 0$. For suitable values of the network parameters,
the two stationarity conditions may simultaneously be satisfied, leading to
fixed points of the learning dynamics, one of them stable and of poor
generalization, the other with one attractive and one repulsive direction.
 Correspondingly, the flow is 
divided by a separatrix defined by a critical $\lambda_c({\cal Q}_0)$
into trajectories leading to convergence according to the asymptotic 
behaviour (\ref{e.easa1})-(\ref{e.qas}) for  $\lambda > \lambda_c({\cal Q}_0)$,
or to poor generalization otherwise.

The salient features of these results for the case of non-structured data
are the convergence of the AR-Hebb-algorithm in 
the sense of {\em asymptotically perfect generalization\/}
with a power law depending on the learning parameters $L$ and $\lambda$
and the existence of a minimal value  $\lambda_c({\cal Q}_0)$, fixed
by the pre-asymptotic structure and below which the system is driven
toward complete confusion.
Notice also that the best convergence is achieved for $\lambda$ just above
$\lambda_c$.  
One last point concerns the recalling parameter $w$, eqs. 
(\ref{e.4}),(\ref{e.r}). 
A rough first quantitative characterization
of this modification would be that it leads to an effective rescaling of the 
parameter $\lambda$, viz. $\lambda \to \lambda/w$, leading to a corresponding 
reduction of critical $\lambda$'s by approximately a factor $p$. This is well 
supported by numerical simulations (see also  Fig. \ref{f.mcsd}
for the case of structured data) and we conclude that the
algorithm is stable against this supplementary
 element of indeterminism.

\subsection{Structured data}

Numerical simulations indicate that for $m \ne 0$ and $0 < |\eta| < 1$
the behaviour of the algorithm for all $w$ is more involved: generically,
no convergence is found in this case for fixed values 
of the learning parameters. This agrees with
the expectations, since, on the one hand
 the situation found at $L=1$, $w=1$ for structured data
\cite{biri} could be expected to hold
 for every $L$, namely that Hebb updating leads to a
nonzero asymptotic 
generalization error. On the other hand, the situation found before
for non-structured data should hold also for  structured data, 
namely that the perceptron rule ($\lambda=0$) (which for $L=1$
was shown to lead to convergence also in the 
structured data case \cite{biri}) does not work
for $L > 1$.

\begin{figure}[htb] 
\vspace{11cm} 
\includegraphics{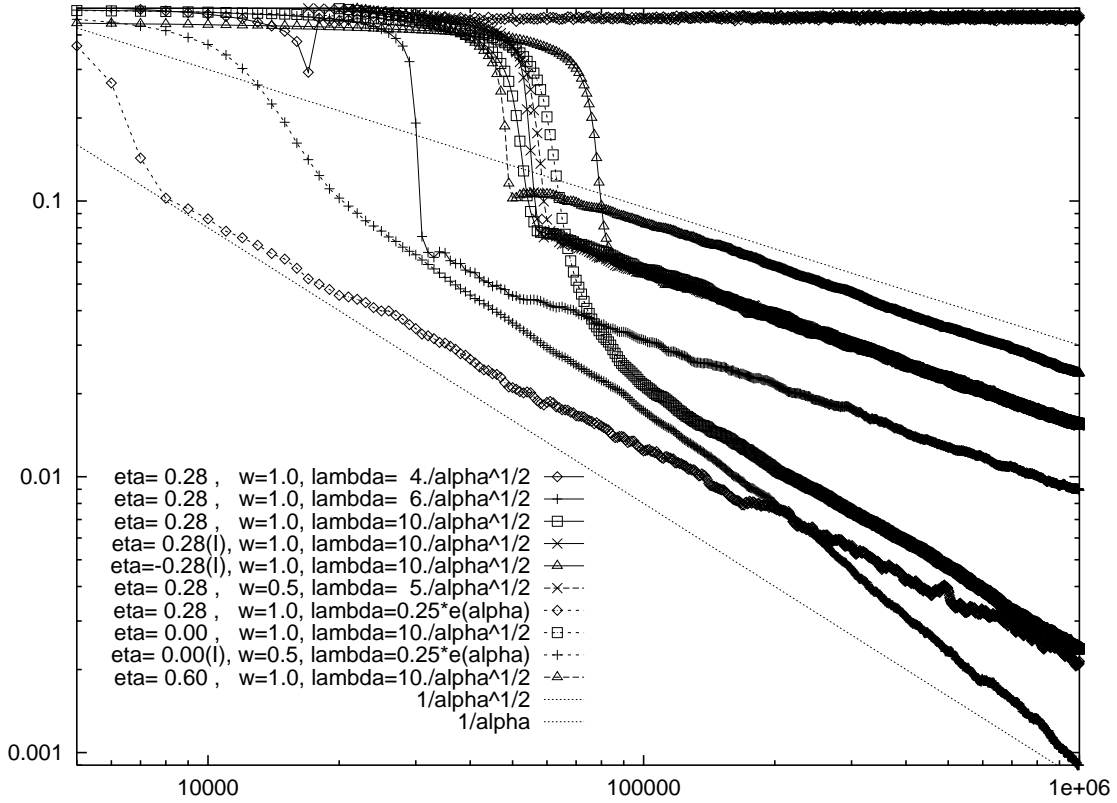} 
\caption{{\it Generalization error $\epsilon_g$ vs. $q =\alpha N/L$ for $L=10$,
$N=100$, 
various overlaps $\eta$ and starting point $\sqrt{{\cal Q}(0)}=100$. We use either patterns
with $\xi_i$: real numbers  or $\xi_i$: Ising spins (I) and
$\lambda=\lambda_0/\sqrt{\alpha}$ or  $\lambda= \lambda_0 \, 
e(\alpha)/e(1)$, see (\ref{e.eqav}). $w$ is the recall
probability (\ref{e.4}),(\ref{e.r}). Note the change in behaviour between
$\lambda_0 = 4$ and $\lambda_0 = 6$. The straight 
lines are illustrative power laws. }}
\label{f.mcsd} 
\end{figure} 
\par\bigskip

In fact one can make a more general argument that for fixed
$\lambda$ the AR-Hebb rule does not lead to perfect generalization for
generically structured data.
To obtain good generalization requires ${\cal R}/\sqrt{\cal Q}\to 1$ and ${\cal D}/
\sqrt{\cal 
Q}\to \eta$, from which one may obtain the necessary dominant scaling (with ${\cal Q}$)
of the various integrals appearing in 
(\ref{e.ajt})-(\ref{e.erf}), namely
\bea
A_{JT} &\simeq& \kappa\ , \nonumber\\
A_{TJ} &\simeq& \sqrt{Q}\, \kappa + o (\sqrt{Q})\ ,\nonumber\\
A_{JJ} &\simeq& \sqrt{Q}\, \kappa + o (\sqrt{Q})\ ,\nonumber\\
A_{JC} &\simeq& \kappa\ , \nonumber\\
A_{TC} &\simeq& \kappa\ , \nonumber\\
A_{TT} &\simeq& \kappa\ , \nonumber\\
S_{JT} &\simeq& 1\ , \nonumber
\eea
\no with
\be
\kappa = m\,\eta\,\varphi(m\,\eta)+ \sqrt{\frac{2}{\pi}}\,\e^{-m^2\,\eta^2/2}\ .
\ee
\no This in 
turn would lead to the following asymptotic expressions for the flow 
equations (\ref{e.dr})-(\ref{e.dq2}) (at fixed $\lambda$)
\bea
\frac{d {\cal R}}{d\alpha} &\simeq& \kappa\, \lambda, \nonumber\\
\frac{d {\cal D}}{d\alpha} &\simeq& \kappa\,\lambda, \nonumber\\
\frac{d {\cal Q}}{d\alpha} &\simeq& \sqrt{\cal Q}\,\kappa\,\lambda + \lambda^2\ . \nonumber
\eea
The solution at large $\alpha$ would be $R \simeq \kappa\, \lambda\, \alpha + R_0$ 
and ${\cal D} \simeq \kappa\, \lambda\, \alpha + D_0$, while ${\cal Q}$ is asymptotically given 
through the implicit equation
\be
\sqrt{\cal Q} \simeq \frac{1}{2}\, \kappa\, \lambda\, \alpha +
\frac{\lambda}{\kappa}\, \ln(\sqrt{\cal Q}\, \kappa + \lambda) + \frac{1}{2}\, \kappa\, \lambda\, 
\kappa_0\ .
\ee
Here $R_0$, $D_0$, and $\kappa_0$ are integration constants. Hence, asymptotically, 
$\sqrt{\cal Q} \sim \frac{1}{2}\, \kappa\, \lambda\, \alpha$ which is {\em incompatible\/}
with the requirement of good generalization ${\cal R}/\sqrt{\cal Q} \to 1$. 
Thus the algorithm will 
{\em not\/} converge, if $\lambda$ is kept fixed. 

The question arises, however, whether a simple extension of the
algorithm may not overcome the Odyssean dilemma hinted at in the 
beginning of this section.
We hence suggest to tune the parameter
$\lambda$ such that it is large enough at small $\alpha$ to 
overcome the pre-asymptotic conditions and it tends to zero at
large $\alpha$ in order to approach asymptotically the perceptron 
rule. As can be seen on Fig. \ref{f.mcsd} this procedure seems successful.

Since the situation is now much more complicated 
we shall not try to solve the general asymptotic problem, as we did in the case of
non-structured data, but we shall limit ourselves to prove that robust
solutions exist.
For this we start with the following ansatz:
\bea
\lambda &=& \lambda_0 \,\alpha^{-r}, \label{e.la}\\
{\cal Q} &=& c^2 \,\alpha^{2q}, \label{e.qa}\\
\epsilon_g &=& a\, \alpha^{-p}, \label{e.ea}\\
\omega &=& b \,\alpha^{-s}. \label{e.oa}
\eea
\no with $\omega$ defined via (\ref{e.con}).
The asymptotic equations  obtained from the flow equations (\ref{e.dr})-(\ref{e.dq2}) 
assuming $ p \sim q \sim r > s \geq 0$  are of the form:
\bea
2\sqrt{{\cal Q}}\,\frac{d\epsilon_g}{d\alpha} &\simeq& 
 A_{11}\,\lambda + \frac{A_{12}}{\sqrt{\cal Q}} 
+A_2\,\epsilon_g, \label{e.dax}\\ 
2\sqrt{{\cal Q}}\sin \frac{z}{2}\,\epsilon_g\,\frac{d\omega}{d\alpha} &\simeq& 
 B_{11}\,\lambda  + 
\frac{B_{12}}{\sqrt{\cal Q}} 
+B_2\,\epsilon_g, \label{e.dao}\\
\frac{d\sqrt{\cal Q}}{d\alpha} &\simeq& C_0\,\lambda +C_1\, \epsilon_g , \label{e.daq}
\eea 
 \no Here the coefficients
$A_{\gamma},B_{\gamma},C_{\gamma}$ are function of $m,z,L$ and of $\omega$ 
(the explicit expressions obtained by Maple are given in Appendix, section
\ref{app2}).

\begin{figure}[htb] 
\vspace{10cm} 
\includegraphics{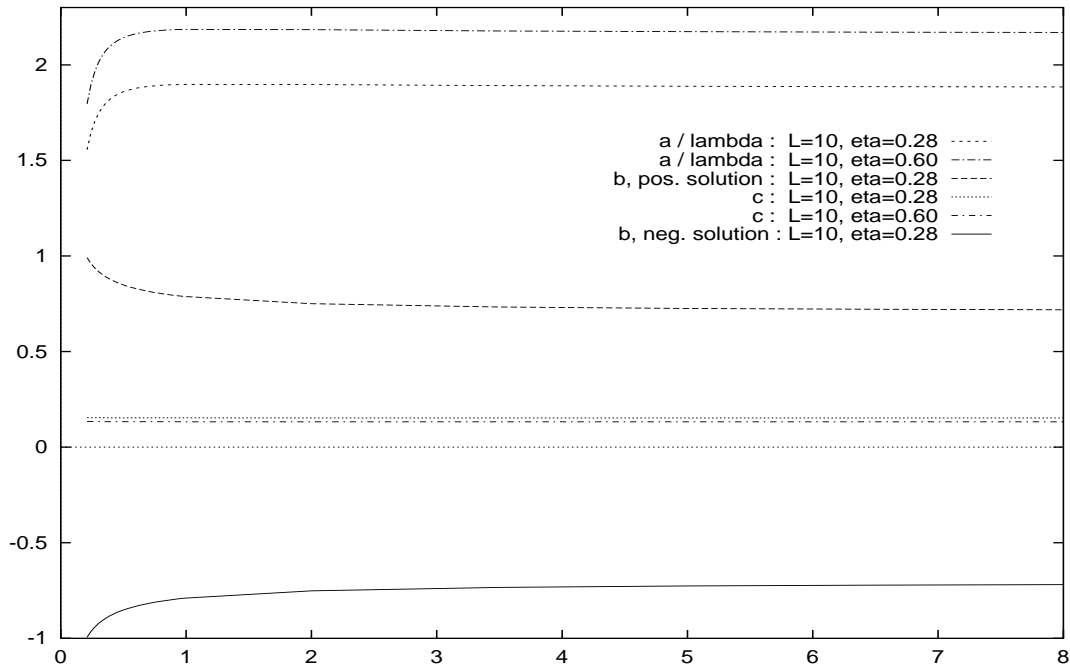} 
\caption{{\it Asymptotic solution (\ref{e.la})-(\ref{e.sola}) for $L=10$:
$a/\lambda_0$, $b$ and $c$ as function of $\lambda_0$ for two 
values of the overlap, $\eta = 0.28,\, 0.6$. There are generally 
two solutions  $a,\, \pm b,\,c$ with $b$ practically independent on $\eta$.
Note that there is no solution for $\lambda_{0} < \lambda_{0,c}^{asympt.} 
\sim 0.2$. }}
\label{f.assol} 
\end{figure}

It is easy to see that an asymptotic  solution can exist for:
\be 
p=q=r=1/2, \ \ s=0 ,\label{e.sola}
\ee
\no which is therefore compatible with the assumptions used to derive the asymptotic
equations (\ref{e.dax})-(\ref{e.daq}). Then $a,b,c$ are obtained as
function of $m,\eta,L$ for given $\lambda_0$, 
with some restrictions on the latter (notice that  the coefficients 
$A_{\gamma},B_{\gamma},C_{\gamma}$ depend 
nonlinearly on $\omega$, hence on $b$). For illustration, we
show in Fig. \ref{f.assol} the values of $a,b$ and $c$ as function of $\lambda_0$
for $L=10,\, m=1$ and 
two values of the data-teacher 
overlap $\eta$. Notice that there is no asymptotic
solution for $\lambda_0$ below $\simeq 0.2$. 

\par\bigskip

\begin{figure}[htb] 
\vspace{12cm} 
\includegraphics{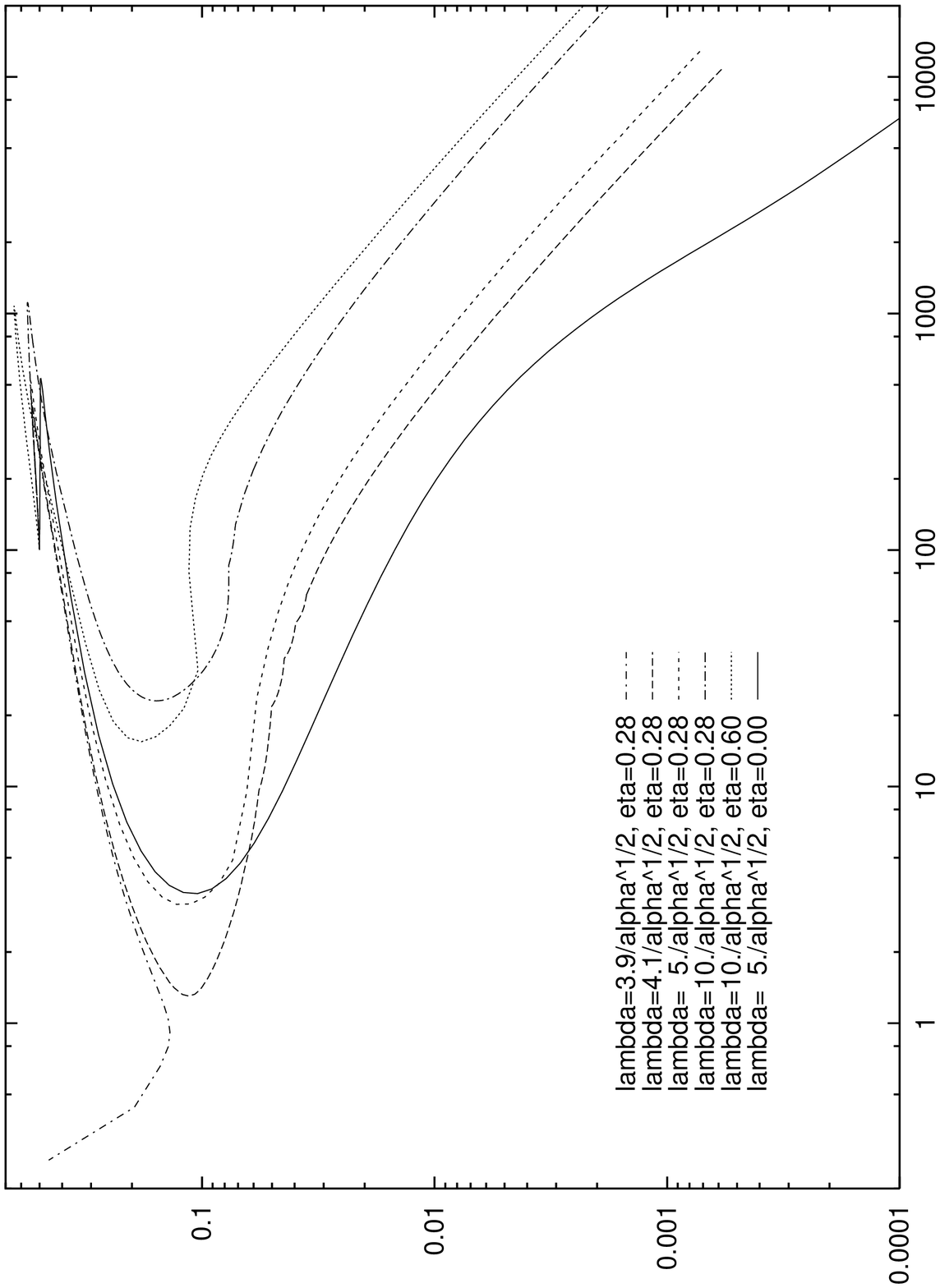} 
\includegraphics{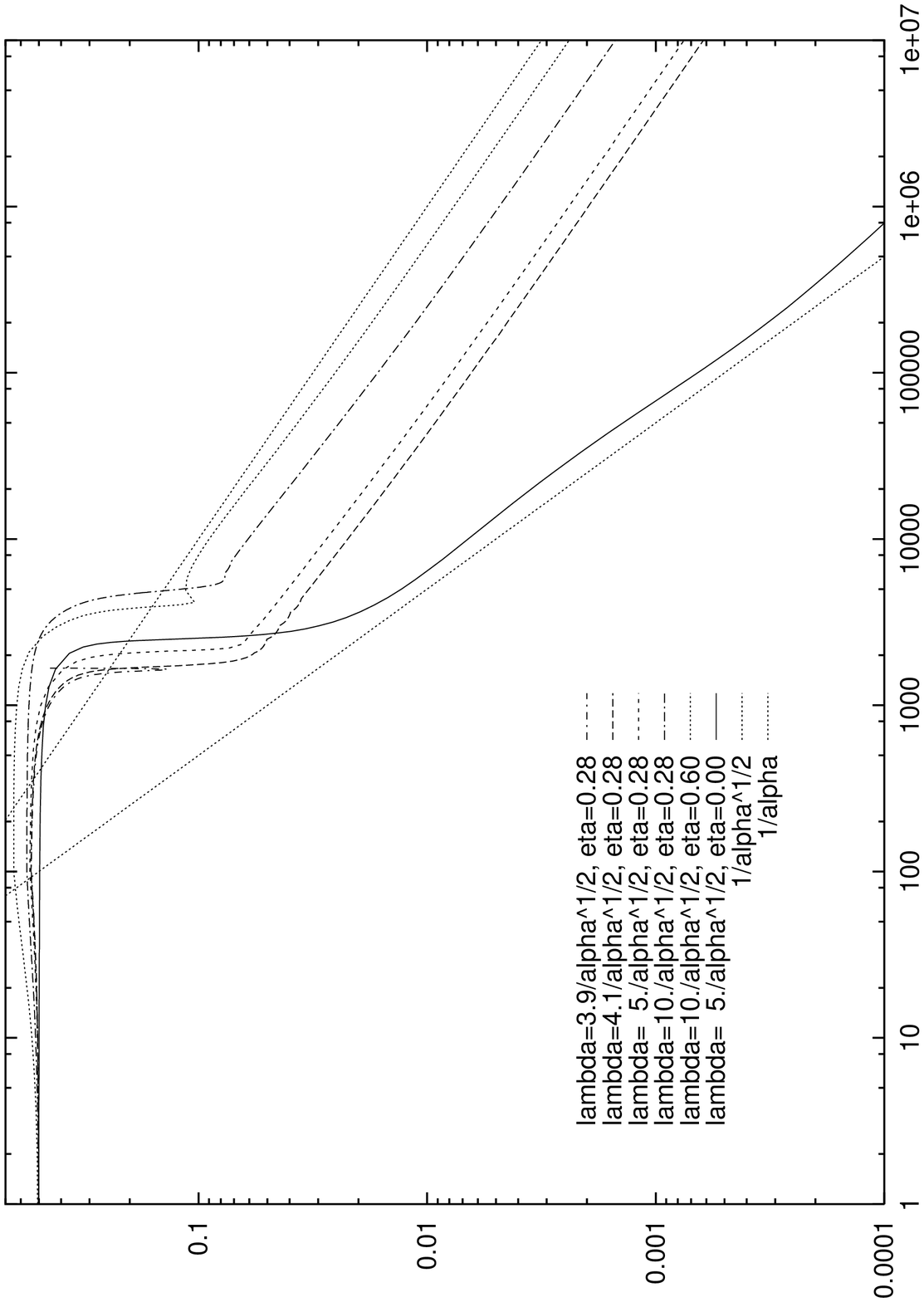} 
\includegraphics{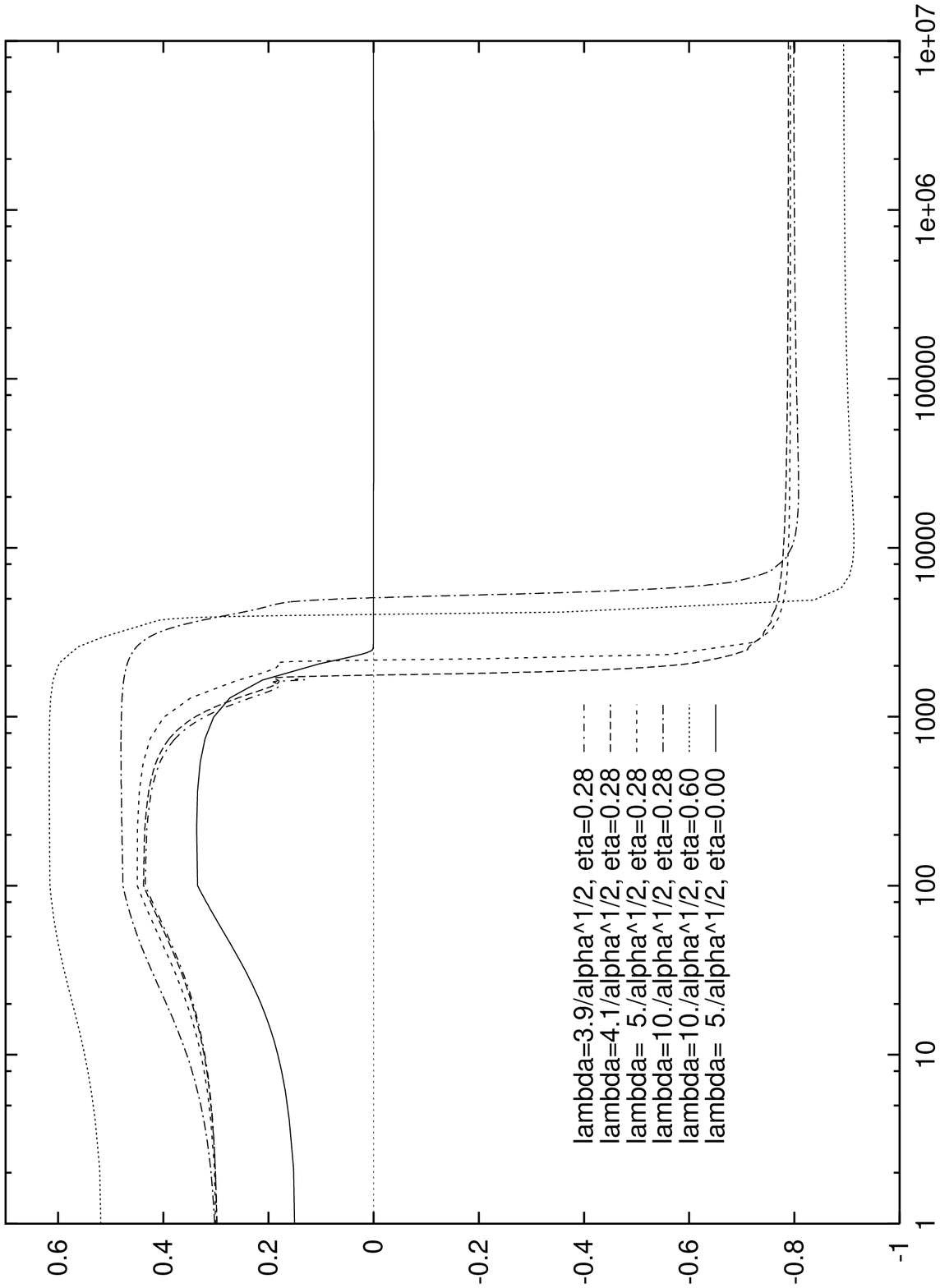} 
\includegraphics{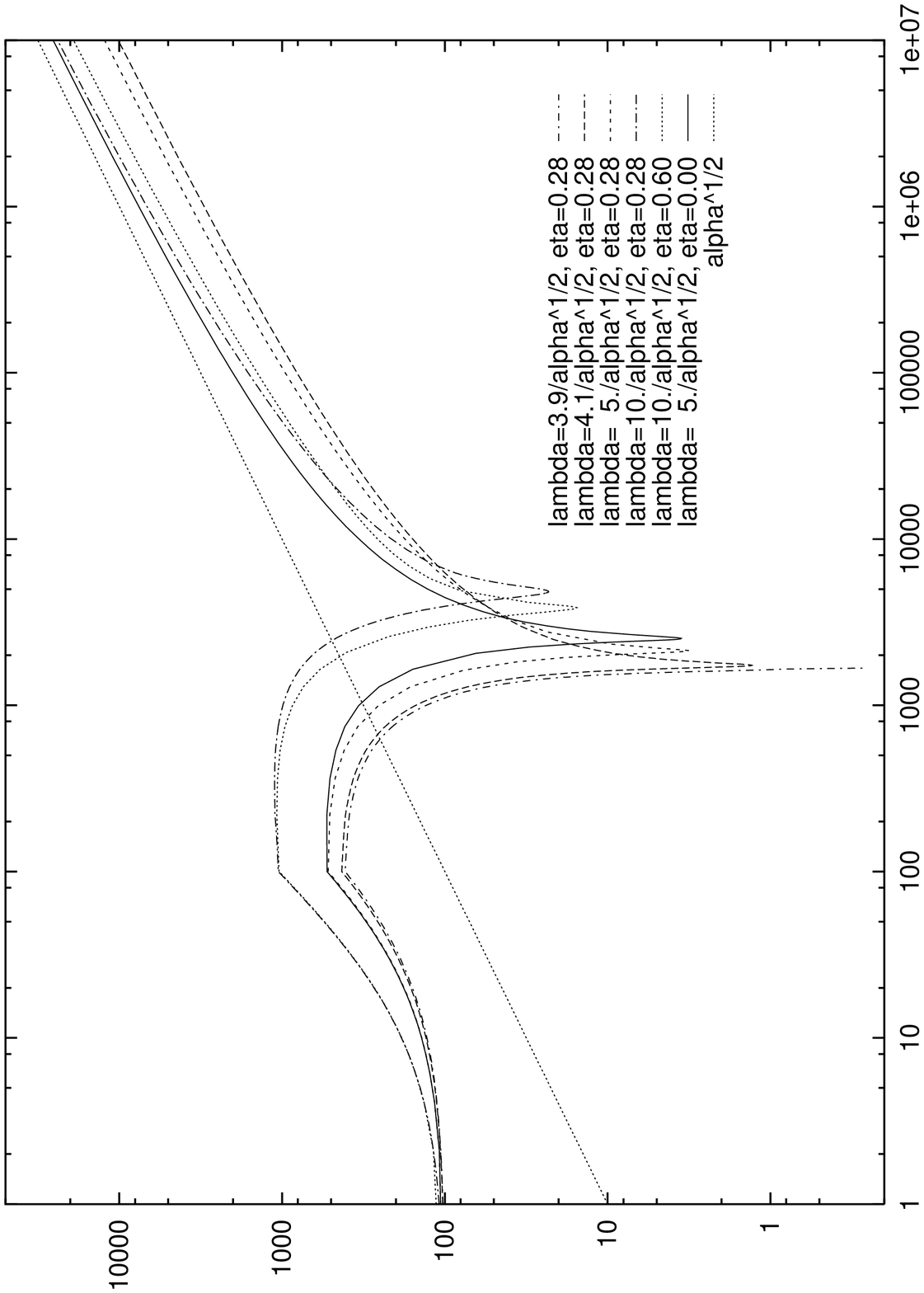} 
\caption{{\it Solution of the flow equations (\ref{e.dr})-(\ref{e.dq2})
for $L=10$, starting point $\sqrt{{\cal Q}(0)} = 100$ and various
overlaps $\eta$:
flow with $\alpha$ in the $\epsilon_g$-$\sqrt{\cal Q}$ plane (upper left),
$\epsilon_g$ vs $\alpha$ (upper right), $\omega$ vs $\alpha$ 
(lower left) and $\sqrt{\cal Q}$ vs $\alpha$ (lower right). 
Note the change in behaviour between
$\lambda_0 = 3.9$ and $\lambda_0 = 4.1$, compare with Fig. \ref{f.mcsd}.
}}
\label{f.sdsol} 
\end{figure}

In Fig. \ref{f.sdsol} we show the solution of the full equations
(\ref{e.dr})-(\ref{e.dq2})  --
compare also with Fig. \ref{f.mcsd} -- which can be seen
to approach the asymptotic solution (\ref{e.la})-(\ref{e.sola}). The solutions are robust 
in the sense that  for all $m,\eta,L$ there exists  a large
region of $\lambda_0$ leading to convergence according to (\ref{e.sola}).
Notice, however, that in the pre-asymptotic region  similar 
phenomena to the non-structured data case seem to take place:
the flow is divided by a separatrix defined by a  $\lambda_{0,c}$
(the MC simulation presents the same effect, see Fig. \ref{f.mcsd}).

\par\bigskip

We have thus shown that the simple decrease of 
$\lambda$ as $1/\sqrt{\alpha}$ provides convergence to asymptotic
perfect generalization with the power $-1/2$. Alternatively, one can
decrease $\lambda$ as $1/\sqrt{\cal Q}$, or as $e(\alpha)$,
where
\be
e(\alpha)=
\frac{1}{\alpha}\sum_{q=1}^{\alpha N/L} e_q
\label{e.eqav}
\ee
\no 
using the running 
``observed error" $e_q$ (\ref{e.2}) (this is in a sense 
the most natural choice, since the student
only applies its observations). 
Again the algorithm is stable 
against noise or a further dilution of the information
introduced by taking $w < 1$. See Fig. \ref{f.mcsd}.

\section{Summary and Discussion}

In the present paper we have investigated the performance of the
AR-Hebb-algorithm introduced in \cite{renu99} in the case where
the input patterns are structured. The pattern statistics is 
characterized by the anisotropy vector $m {\vec C}$ and performance
of the learning rule depends on $m$ and on the overlap $\eta$ between
the anisotropy vector and the vector ${\vec B}$ that defines the rule 
-- apart from the parameters $\lambda$ and $L$ which characterize the
AR-Hebb-algorithm.

As for usual Hebb learning, a tuning of learning parameters is {\em
required\/}
to achieve good generalization for the classification of structured
patterns. 
Given $L$, the only free parameter of the
algorithm is $\lambda$, and tuning of $\lambda$ may proceed in various
ways. For instance, one may scale $\lambda$ either with $\alpha$, i.\,e.,
with
the number of input-output pairs presented, or with the self-overlap
${\cal Q}$,
or with the empirical error-rate $e_q$. Our analysis reveals that the
scaling 
$\lambda \sim \alpha^{-1/2}$, which according to that analysis is
equivalent 
to the scalings  $\lambda \sim {\cal Q}^{-1/2}$, or 
$\lambda\sim e(\alpha)$, leads
to 
asymptotically perfect generalization. The behaviour is robust in the
sense 
that the prefactor $\lambda_0$ may be varied over a wide range without
changing
the asymptotic scaling of the generalization error. In this sense the
tuning
required to obtain a working algorithm is not fine-tuning. The only 
requirement for obtaining good generalization is that $\lambda_0$ in 
(\ref{e.la}) exceeds a certain minimum value, $\lambda_{0,c}^{asympt.}$. This behaviour is
reminiscent 
of the fact that a minimum value of $\lambda$ was also required in the
case 
of unstructured data. In that case, however, the reason was entirely
related 
to pre-asymptotic behaviour related to the fixed-point structure of the
flow 
equations, whereas the above analysis is restricted to the asymptotic
domain.

From the numerical solution of the full flow-equations and from
simulations,
we see some empirical evidence that a non-trivial fixed point structure 
governing the pre-asymptotic behaviour in analogy to what has been found
in
\cite{renu99} is present also in the case studied here.
As the 
present dynamical problem is {\em three}-dimensional instead
two-dimensional,
however, the consequences of this might be suspected to be less severe. For
instance, 
a fixed-point with stable and unstable directions in three dimensions does
not 
necessarily produce a separatrix as in the two-dimensional case.
However, the projection onto the $\epsilon_g$-$\sqrt{\cal Q}$ plane shows
a separatrix  and hence a $\lambda_{0,c}$, as in the unstructured
data case 
(see Fig. \ref{f.sdsol}), with $\lambda_{0,c} > \lambda_{0,c}^{asympt.}$.
Unlike in 
the two-dimensional case with unstructured patterns, we have so far {\em
not\/}
found any evidence of non-universal behaviour of the generalization curve. 
Whether this is intrinsically related to the different role fixed points
appear 
to play in the present case, we do at present not know. Exceptional 
behaviour appears for 
 $\eta=0$ which in the student-teacher scenario, however, 
is equivalent to the unstructured case.\par\bigskip

{\bf Acknowledgments}:
This project was initiated during the Seminar 
`Statistical Physics of Neural Networks' in Dresden, March
1999.  The authors would like to thank the Max Planck Institut
f\"ur Physik Komplexer Systeme in Dresden for hospitality 
and financial support and the participants to the
workshop for interesting discussions.

\section{Appendix}

\subsection{Expectations values} \label{app1}

The expectations values $A_.,S_.$ in (\ref{e.dr})-(\ref{e.dq2}) are:
\bea
A_{JT} &=& m\,\eta\,\varphi(m\,\frac{\cal D}{\sqrt{\cal Q}})+
\sqrt{\frac{2}{\pi}}\,\frac{\cal R}{\sqrt{\cal Q}}\,
\e^{-\frac{m^2{\cal D}^2}{2{\cal Q}}} 
 \nonumber \\
&=& m\,\cos z \,\varphi(m\cos y)+\sqrt{\frac{2}{\pi}}\,\cos x \,
\e^{-\frac{1}{2}m^2\cos^2y}, \label{e.ajt}\\
A_{TT} &=& m\,\eta\,\varphi(m\,\eta)+
\sqrt{\frac{2}{\pi}}\,
\e^{-\frac{m^2{\eta}^2}{2}}
 \nonumber \\
&=&  m\, \cos z \,\varphi(m\cos z)+\sqrt{\frac{2}{\pi}}\  
\e^{-\frac{1}{2}m^2\cos^2z}, \label{e.att}\\
S_{JT} &=& 1 + \varphi(m\,\frac{\cal D}{\sqrt{\cal Q}}) - 
\varphi(m\,\eta) - 
4\,G(\frac{\cal R}{\sqrt{\cal Q}},\frac{\cal D}{\sqrt{\cal Q}},\eta)
 \nonumber \\
&=& 1 + \varphi(m\cos y) - \varphi(m\cos z) - 4\,G(\cos x,\cos y,\cos z), 
\label{e.sjt}\\
A_{JJ} &=& m\,\frac{\cal D}{\sqrt{\cal Q}}\,
\varphi(m\,\frac{\cal D}{\sqrt{\cal Q}})+
\sqrt{\frac{2}{\pi}}\,
\e^{-\frac{m^2{\cal D}^2}{2{\cal Q}}}
 \nonumber \\
&=&  m\, \cos y\, \varphi(m\cos y)+\sqrt{\frac{2}{\pi}}\, 
\e^{-\frac{1}{2}m^2\cos^2y}, \label{e.ajj}\\
A_{TJ} &=& m\,\frac{\cal D}{\sqrt{\cal Q}}\,\varphi(m\,\eta)+
\sqrt{\frac{2}{\pi}}\,\frac{\cal R}{\sqrt{\cal Q}}\,
\e^{-\frac{m^2{\eta}^2}{2}}
 \nonumber \\
&=& m\, \cos y\, \varphi(m\cos z)+\sqrt{\frac{2}{\pi}}\,\cos x \,
\e^{-\frac{1}{2}m^2\cos^2z}, \label{e.atj}\\
A_{JC} &=& m\,\varphi(m\,\frac{\cal D}{\sqrt{\cal Q}})+\sqrt{\frac{2}{\pi}}
\,\frac{\cal D}{\sqrt{\cal Q}}\,
\e^{-\frac{m^2{\cal D}^2}{2{\cal Q}}}
 \nonumber \\
&=& m\, \varphi(m\cos y)+\sqrt{\frac{2}{\pi}}\,\cos y \,
\e^{-\frac{1}{2}m^2\cos^2y}, \label{e.ajc}\\
A_{TC} &=& m\,\varphi(m\,\eta)+
\sqrt{\frac{2}{\pi}}\,\eta\,
\e^{-\frac{m^2{\eta}^2}{2}} 
\nonumber \\
&=&  m\,  \varphi(m\cos z)+\sqrt{\frac{2}{\pi}}\,\cos z \,
\e^{-\frac{1}{2}m^2\cos^2z}, \label{e.atc}
\eea
\no where 
\bea
G(\frac{\cal R}{\sqrt{\cal Q}},\frac{\cal D}{\sqrt{\cal Q}},\eta) &=& \frac{1}{2}\int_{-\infty}^{m\,\frac{\cal D}{\sqrt{\cal Q}}}
\frac{dt}{\sqrt{2\pi}}\,\e^{-\frac{1}{2}t^2}\left(1+
\varphi\left(\frac{t\,\frac{\cal R}{\sqrt{\cal Q}} -m\,\eta}
{\sqrt{1-\frac{{\cal R}^2}{\cal Q}}}\right)\right) \nonumber\\
&=& \frac{1}{2}\int_{-\infty}^{m\,\cos y}\frac{dt}{\sqrt{2\pi}}\,
\e^{-\frac{1}{2}t^2}\left(1+\varphi\left(\frac{t\,\cos x -m\,\cos z)}{\sin x}\right)\right)
\label{e.int}\\
\varphi(x) &=& {\rm erf}(x/\sqrt{2}) \label{e.erf}
\eea
\no with erf the error function.

\subsection{Asymptotic coefficients} \label{app2}

The Maple expressions for the coefficients $A_{\gamma},B_{\gamma},C_{\gamma}$ in 
(\ref{e.dax})-(\ref{e.daq}) are:

\bea
u&=&\frac {1}{2} \,m\,{\rm cos}(z)\,\sqrt{2}\\
v&=&m\,{\rm sin}({ \frac {1}{2}} \,z)\,\sqrt{2}\\
g(v \omega)&=& v\, \omega\,{\rm erf}(v\, \omega)\\
f(v \omega)&=& v\, \omega\,{\rm erf}(v\, \omega) + 
{\displaystyle \frac {e^{ - v^{2}\,
 \omega^{2}}}{\sqrt{\pi }}}\\
A_{11} &=&  - 4\,{\displaystyle \frac {{\rm sin}({ \frac {1}{2}} \,z)
\,{\rm erf}(u)\, \omega}{\pi 
}} \\ 
 A_{12} &=& {\displaystyle \frac {e^{ - u^{2}}\, f(v \,\omega)
}{\pi \sqrt{\pi}\,L}} 
\\
 A_{2} &=& {\displaystyle \frac {e^{ - u^{2}}\,( - 2
\,{\displaystyle \frac {1 + 2\,v^{2}\, \omega^{2}}{L}}  + 4\,(1 - 
{\displaystyle \frac {1}{L}} )\,{\rm erf}(u)\,f(v\,\omega))}{\sqrt{2\pi }}} 
\\
 B_{11} &=&   4\,{\displaystyle \frac {m\,{\rm sin}({ \frac {1}{2}} \,z)^{2}
\,{\rm erf}(u)\,
(\omega^{2}-1+s^2)}{\pi }} 
\\
 B_{12} &=&  - {\displaystyle \frac {{\rm sin}({ \frac {1}{2}} \,z)
\,e^{ - u^{2}}\, \omega\,f(v\,\omega)}{
\pi ^{3/2}\,L}} 
\\
 B_{2} &=& 4\,{\displaystyle \frac {m\,{\rm sin}({ \frac {1}{2}} \,z)^{2}
\,e^{ - u^{2}}\,( \omega
^{2} - 1 + s^2)\,({\displaystyle \frac {v\, \omega}{L}}  - (1
 - {\displaystyle \frac {1}{L}} )\,{\rm erf}(u)\,f(v\,\omega))}{\sqrt{\pi }}
} 
\\
 C_{0} &=& \sqrt{2}\,u\,{\rm erf}(u) + {\displaystyle \frac {
\sqrt{2}\,e^{ - u^{2}}}{\sqrt{\pi }}} 
\\
 C_{1} &=&  - (1 - {\displaystyle \frac {1}{L}} )\,e^{ - u^{
2}}\,f(v\,\omega)\,(\sqrt{\pi }\,m\,{\rm erf}(u)\,{\rm cos}(z) + \sqrt{2}\,
e^{ - u^{2}})
\eea

\end{document}